\documentclass[prb,manuscript]{revtex4}
\usepackage{amssymb,amsmath,bm}
\font\bba=msbm10 scaled 1080
\font\bbb=msbm8 
\font\bbc=msbm6 
\newfam\bbfam
\textfont\bbfam=\bba
\scriptfont\bbfam=\bbb
\scriptscriptfont\bbfam=\bbc
\def\bb{\fam\bbfam\bba}
\def\R{{\bb R}}

\begin{document}
\title{The Diffusion Equation on a Hypersphere}
\author{Jean-Michel Caillol}
\affiliation{Laboratoire de Physique Th\'eorique \\
UMR 8267, B\^at. 210 \\
Universit\'e de Paris-Sud \\
91405 Orsay Cedex, France}
\email{Jean-Michel.Caillol@th.u-psud.fr}
\date{\today}
\begin{abstract}
We study the diffusion equation  on the surface of a $4D$ sphere and obtain 
a Kubo formula for the diffusion coefficient.   
\end{abstract}
\maketitle
\newpage
Since  numerical simulations of fluids or plasmas within
hyperspherical boundary conditions are appealing, it is not without interest
to dispose of an explicit expression for the diffusion coefficient
$D$ in terms of time averages of microscopic quantities.\cite{Hansen} 
Our starting point will be the diffusion equation 

\begin{equation}
\label{diff}
\mathcal{D} \rho(M,t) \equiv\left(\frac{\partial}{\partial t} -D\Delta^{\mathcal{S}_3} \right)
 \rho(M,t)=0 \; ,
\end{equation}
where the point $M$ lives on the surface of the
four dimensional ($4D$) sphere $\mathcal{S}_3$ of center $O$ and radius $R$
(a hypersphere for short). $D (>0)$ is the diffusion coefficient 
and $\Delta^{\mathcal{S}_3}$ is the 
Laplacian appropriate to the geometry.\cite{Nissfolk,Caillol1} 
Eq.\ (\ref{diff}) was solved recently by Nissfolk \textit{et al.}; their
expression for $\rho(M,t)$ is however akward and does not allow to obtain an
explicit expression for $D$. A more convenient expression  for $\rho(M,t)$
is obtained in this paper which yields an explicit expression for $D$.

We first note that the Brownian motion of a
particle in $\mathcal{S}_3$ may also be viewed as the  Brownian rotation of a
linear $4D$ molecule. As the expression of the rotational diffusion coefficient
of a $3D$ rotator is known since the work of Debye \cite{Debye,Berne,Caillol2}
it should be easy to extend this old result to the $4D$ case.
Following Berne and Pecora \cite{Berne} we first expand $\rho(M,t)$ 
on the complete basis set of hyperspherical
harmonics \cite{Caillol1,Bander}
\begin{equation}
\label{harm}
\rho(M,t)=\sum_{L=0}^{\infty} \sum_{\bm{\alpha}} \rho_{L,\bm{\alpha}}(t) \;
 Y_{L,\bm{\alpha}}(\bm{\xi}) \; ,
\end{equation}
where the unit vector $\bm{\xi}=\bm{OM}/R$ denotes the orientation of vector
$\bm{OM}$ in $\R^4$ 
and $Y_{L,\bm{\alpha}}(\bm{\xi})$ is a $4D$ hyperspherical harmonics.
We recall
that the quantum number $L$ is a positive integer and that $\bm{\alpha}=(m_1,m_2)$ where
$m_i \; (i=1,2)$ takes the $L+1$ values $m_i= -L/2,-L/2+1,\ldots,
L/2$. 
The $4D$ harmonics $Y_{L,\bm{\alpha}}(\bm{\xi})$ satisfy the following
properties:\cite{Caillol1,Bander}
\begin{itemize}
\item{(i)} orthogonality
\begin{equation}
\label{ortho}
\int d\Omega(\bm{\xi})\; Y_{L,\bm{\alpha}}^{\star}(\bm{\xi})
 Y_{L',\bm{\alpha}'}(\bm{\xi})=\delta_{L,L'} \;
 \delta_{\bm{\alpha},\bm{\alpha}'}
 \; ,
\end{equation}
where $ d\Omega(\bm{\xi})$ denotes the infinitesimal volume element 
(or $4D$ infinitesimal solid angle) on the
hypersphere of unit radius.  
\item{(ii)} completeness
\begin{equation}
\label{compl}
\sum_{L,\bm{\alpha}} Y_{L,\bm{\alpha}}^{\star}(\bm{\xi})
 Y_{L,\bm{\alpha}}(\bm{\xi}')=\delta^{\mathcal{S}_3}(\bm{\xi},\bm{\xi}') \; ,
\end{equation}
where
$ \delta^{\mathcal{S}_3}$ is the Dirac distribution for the unit hypersphere
defined as
\begin{equation}
\int d\Omega(\bm{\xi})\; f(\bm{\xi}) \delta^{\mathcal{S}_3}(\bm{\xi},\bm{\xi}')
= f(\bm{\xi}'). 
\end{equation}
\item{(iii)} addition theorem
\begin{equation}
\label{add}
\sum_{\bm{\alpha}} Y_{L,\bm{\alpha}}^{\star}(\bm{\xi})
 Y_{L,\bm{\alpha}}(\bm{\xi}')=P_{L}(\bm{\xi}\cdot \bm{\xi}') \; ,
\end{equation}
where the dot denotes the usual $4D$ scalar product, i.e 
$\bm{\xi}\cdot \bm{\xi}'=\cos(\psi)$ where $\psi$ is the angle between the two
unit vectors $\bm{\xi}$ and $\bm{\xi}'$ and 
\begin{equation}
\label{poly}
P_{L}\left(\cos \left(\psi \right) \right)= \frac{L+1}{2 \; \pi^2} 
\frac{\sin (L+1)\psi}{\sin(\psi)}\; . 
\end{equation}
Note that $R\psi$ is the geodesic length between the points $M$ and $M'$ of
$ \mathcal{S}_3$.
\end{itemize}
Moreover the $Y_{L,\bm{\alpha}}(\bm{\xi})$  are the
eigenvectors of the Laplace-Beltrami operator $\Delta^{\mathcal{S}_3}$ with
eigenvalues $-L(L+2)/R^2$. \cite{Caillol1,Bander}It follows from this important
property
that eq.\ (\ref{diff}) is equivalent to the set of equations
\begin{equation}
\label{diff2}
\left(\frac{\partial}{\partial t} +D \; \frac{L(L+2)}{R^2} \right)
 \rho_{L,\bm{\alpha}}(t)=0 \; ,
\end{equation}
the solution of which reads obviously as 
\begin{equation}
\label{solu}
\rho_{L,\bm{\alpha}}(t)=
\rho_{L,\bm{\alpha}}(0) \exp\left(-D L\left(L+2\right)t/R^2\right) \; .
\end{equation}

We shall denote by $\rho(M,t\vert M_0,0)$ the solution of\ (\ref{diff})
corresponding to
the initial condition $\rho(M,0)=\delta^{\mathcal{S}_3}(M,M_0)\equiv R^{-3} \;\delta^{\mathcal{S}_3}(\bm{\xi}
,\bm{\xi}_0)$, i.e.
the solution of 
\begin{equation}
\label{Green}
\mathcal{D}\rho(M,t\vert M_0,0)=\delta(t) \; 
R^{-3} \;\delta^{\mathcal{S}_3}(\bm{\xi}
,\bm{\xi}_0)
\; .
\end{equation}
It follows readily from eqs.\ (\ref{solu}) and\ (\ref{compl}) that the Green
function $\rho(M,t\vert M_0,0)$ can be expressed as
\begin{eqnarray} 
\label{Green-solu}
\rho(M,t\vert M_0,0)&=& 0 \; \; (t<0) \nonumber \\
\rho(M,t\vert M_0,0) &= & 
\frac{1}{R^3} \sum_{L,\bm{\alpha}} Y_{L,\bm{\alpha}}^{\star}(\bm{\xi}_0)
Y_{L,\bm{\alpha}}(\bm{\xi}) \; \exp\left(-D L\left(L+2\right)t/R^2\right) 
\; \; (t>0)
\; .
\end{eqnarray}
Eq.\ (\ref{Green-solu}) can be further simplified with the help of the addition
theorem\ (\ref{add}) which yields our final result
\begin{equation}
\label{Green-solu2}
\rho(M,t\vert M_0,0)=
\frac{1}{R^3} \sum_{L} P_{L}(\bm{\xi}_{0} \cdot \bm{\xi}) \;
\exp\left(-D L\left(L+2\right)t/R^2\right) \; \; (t>0) \; .
\end{equation}
Some comments are in order.

(i) The solution\ (\ref{Green-solu2}) is invariant upon rotations about the axis
$\bm{\xi}_0$ as expected. Eq.\ (\ref{Green-solu2}) is a trivial generalization
 of Debye's
result\cite{Debye,Berne,Caillol2} (for the $3D$ rotor the Tchebychev polynomial
$P_L$ in eq.\ (\ref{Green-solu2}) has to be replaced by a Legendre polynomial, up
to a multiplicative constant)

(ii) Note that 
\begin{equation}
\int R^3 d\Omega(\bm{\xi})\; \rho(M,t\vert M_0,0)= 1   \; \; (t>0) \;,
\end{equation}
i.e. the probability is conserved, the particle does not evaporate.

(iii) We have also
\begin{equation}
\lim_{t \to + \infty} \rho(M,t\vert M_0,0)=\frac{1}{2 \pi^2 R^3} \; ,
\end{equation} 
i.e. the solution of the diffusive process is uniform after infinite time
(the volume of
$\mathcal{S}_3$ is equal precisely to $2 \pi^2 R^3$)

(iv) Let us define, following Berne and Pecora\cite{Berne}, the time correlation
function
\begin{eqnarray}
C_{L}(t) & \equiv &  \langle P_{L}
\left( \bm{\xi}\left(0\right)\cdot \bm{\xi}\left(t\right)\right)
 \rangle \nonumber \\
&=& \int \frac{ d\Omega({\bm{\xi}_0})}{2 \pi^2} \;
\int R^3  d\Omega({\bm{\xi}}) \;  P_{L}(\bm{\xi}_0 \cdot \bm{\xi})
\; \rho(M,t\vert M_0,0) \; ,
\end{eqnarray}
where we have averaged the initial position uniformly on $\mathcal{S}_3$.
As a consequence of the orthogonality properties of the $P_L$ we find that
\begin{equation}
\label{D}
C_{L}(t) =\frac{ (L+1)^2}{2 \pi^2}
 \exp \left(-D L\left(L+2\right)t/R^2\right) \; .
\end{equation} 
Defining now the reorientational time $\tau_L$ of our $4D$ rotor  as 
\begin{eqnarray}
\tau_L &= &\int_0^{\infty} dt \; \frac{C_{L}(t)}{C_{L}(0)} \nonumber \\
       &= & \frac{R^2}{D L (L+2)} \; ,
\end{eqnarray}    
we have now at our disposal an explicit expression of the diffusion coefficient
$D$.  Note the aesthetic relation $\tau_L /\tau_{L'}=L'(L'+2)/L(L+2)$ which
generalizes Debye's relation to the $4D$ rotor. 

(v) It does not seem possible to obtain from eqs.\ (\ref{D}) an expression for $D$
in term of the time autocorelation function of the velocity of the particle. 
Moreover one can deduce from eq.\ (\ref{Green-solu2}) that the mean
square displacement of the stereographic
projection of point $M$ vanishes for $t \to \infty$.

It remains now to show that the expression\ (\ref{Green-solu2}) of 
$\rho(M,t\vert M_0,0)$ is equivalent to that of ref.\onlinecite{Nissfolk}. This
can be done as follows. Let us rewrite the reduced
 $\widehat{\rho}= 2 \pi^2 R^3 \rho(M,t\vert M_0,0)$
as
\begin{equation}
\widehat{\rho}=\sum_{L=0}^{\infty} (L+1) \frac{  \sin(L+1) \psi}{\sin \psi}
\exp(-K \; L (L+2)) \; ,
\end{equation}
where $K=Dt/R^2$. {\em A priori} the angle $\psi$ is in the range $(0,\pi)$,
however since the function is formally even in $\psi$ we define 
$\widehat{\rho}(-\psi)=\widehat{\rho}(\psi)$
for negative angles. This gives us a periodic function of period $2 \pi$ defined
for all $\psi \in \R$. We introduce now the periodic function
\begin{equation}
F(\psi)=\int_{0}^{\psi} d \psi ' \;  \widehat{\rho}(\psi ') \sin(\psi ') 
\end{equation}
which can be rewritten after some algebra as
\begin{equation}
F(\psi)= F_0 -\frac{\exp{K}}{2} \sum_{p= -\infty}^{+\infty}
\exp\left(-K p^2\right) \exp (-i p \psi) 
\end{equation}
where $F_0$ is some unessential constant independent of angle $\psi$.
At this point we  recall Poisson
summation theorem which states that for any function $\varphi(x)$ holomorphic in
the strip $-a< \Im z <a$ one has
\begin{equation}
\label{Poisson}
\sum_{n= -\infty}^{+\infty} \varphi(x+2 n \pi)=
\frac{1}{2\pi} \sum_{p= -\infty}^{+\infty} \exp(-ipx) \; 
\widetilde{\varphi}(p) \; ,
\end{equation}
where 
\begin{equation}
\widetilde{\varphi}(p)=\int_{-\infty}^{+\infty} dx \; \varphi(x) \exp(i p x ) \; .
\end{equation}
Applying Poisson theorem for the Gaussian we get 
\begin{equation}
F(\psi)= F_0 - \frac{\sqrt{\pi}\exp{K}}{2 \sqrt{K}}
\sum_{n=-\infty}^{+\infty} \exp \left(  
-\frac{(\psi + 2n \pi)^2}{4K}\right) \; ,
\end{equation}
which, after differentiation yields for $\widehat{\rho}$ 

\begin{equation}
\widehat{\rho}(\psi,t)= \frac{\sqrt{\pi}\exp{K}}{4K^{3/2} \sin \psi}
\sum_{n=-\infty}^{+\infty} (\psi + 2n \pi)
\exp \left(  
-\frac{(\psi + 2n \pi)^2}{4K}\right)  \; ,
\end{equation}
which coincides with the result of ref.\onlinecite{Nissfolk} apart  the
prefactor which is not specified.
\begin{acknowledgments}
The author likes to thank J.-J. Weis for drawing his attention to
ref.~\onlinecite{Nissfolk}.
\end{acknowledgments}

\end{document}